\documentclass[aps,prl,twocolumn,showpacs,superscriptaddress]{revtex4}
\usepackage{hyperref}
\usepackage{natbib}
\usepackage{graphicx}

\begin{document}

\title{Correlation effects in the electronic structure of Mn$_4$ molecular
magnet}

\author{D. W. Boukhvalov}
\affiliation{Institute for Molecules 
and Materials,
Radboud University of Nijmegen, NL-6525 ED Nijmegen, the Netherlands}
\affiliation{Institute of Metal Physics, Russian Academy of Sciences
  Ural Division, Ekaterinburg 620219, Russia}
\author{L. I. Vergara}
\affiliation{Department of Chemistry, University of Tennessee,
Knoxville, TN 37996-1600}
\author{V. V. Dobrovitski}
\affiliation{Ames Laboratory, Iowa State University, Ames IA 50011, USA}
\author{M. I. Katsnelson}
\affiliation{nstitute for Molecules 
and Materials,
Radboud University of Nijmegen, NL-6525 ED Nijmegen, the Netherlands}
\author{A. I. Lichtenstein}
\affiliation{Institute of Theoretical Physics, University of Hamburg,
20355 Hamburg, Germany}
\author{P. K\"ogerler}
\affiliation{Ames Laboratory, Iowa State University, Ames IA 50011, USA}
\affiliation{Institut f\"ur Anorganische Chemie, 
RWTH Aachen University, D-52074 Aachen, Germany}
\author{J. L. Musfeldt}
 \affiliation{Department of Chemistry,
University of Tennessee, Knoxville, TN 37996-1600}
\author{B. N. Harmon}
\affiliation{Ames Laboratory, Iowa State University, Ames IA 50011, USA}

\date{\today}
\begin{abstract}
We present joint theoretical-experimental study of the
correlation effects in the electronic structure of 
(pyH)$_3$[Mn$_4$O$_3$Cl$_7$(OAc)$_3$]$\cdot$2MeCN 
molecular magnet (Mn$_4$).
Describing the many-body effects by cluster dynamical mean-field
theory, we find that Mn$_4$ is predominantly
Hubbard insulator with strong electron correlations. 
The calculated electron gap (1.8~eV) agrees well with the
results of optical conductivity measurements, while other methods,
which neglect many-body effects or treat them in a simplified
manner, do not provide such an agreement. Strong
electron correlations in Mn$_4$ may have important implications
for possible future applications.
\end{abstract}


\maketitle

Single molecule magnets (SMMs), made of
exchange-coupled magnetic ions surrounded by large organic
ligands, represent a novel interesting class of magnetic
materials. They are of fundamental interest as
test systems for studying magnetism at nanoscale, and
interplay between the structural, electronic, and magnetic
properties. SMMs demonstrate fascinating
mixture of clasiscal and quantum properties:
as classical superparamagnets,
they possess large anisotropy and magnetic moment,
but also exhibit interesting mesoscopic
quantum spin effects \cite{qtm,qtmBerry,Mn4qtm}. 
Moreover, recent experiments on the electron transport 
through SMMs \cite{transportSMM}, 
and predicted connection between the transport
and spin tunneling \cite{RomeikeEtal}, make SMMs good candidates for
interesting spintronics studies.
Progress in this area --- synthesis of novel SMMs
with optimized properties, design and analysis of the transport
experiments, possible uses in information processing
--- demands detailed theoretical
investigations of the magnetic and electronic structure of SMMs
\cite{v15chi,Boukhvalov2007,v15xes,pederson,pedersonMn4}.
Among other factors, the many-body correlations caused by the
Coulomb repulsion between electrons, may be important.
E.g., in transition metal-oxide systems \cite{tmox}, which share many similarities 
with SMMs, strong correlations may form the Mott-Hubbard insulating 
state \cite{mott}, where the nature of the charge
and spin excitations is drastically different from
the predictions of standard band-insulator theory.
This affects the
basic properties of the system (e.g., exchange interactions),
and drastically changes charge and spin transport. 

In this joint experimental-theoretical work, we present a
detailed study of the many-body effects in electronic structure of
SMMs
(pyH)$_3$[Mn$_4$O$_3$Cl$_7$(OAc)$_3$]$\cdot$2MeCN (denoted below
as Mn$_4$ for brevity) \cite{Wang1996}. We use the cluster
LDA+DMFT method \cite{cdmft1} which combines the realistic
{\it ab initio\/} calculations based on the local density
approximation (LDA), and the accurate description of the 
correlation effects within the cluster dynamical mean field theory
(CDMFT). Using the electron gap as a most convenient
benchmark, we show that the gap value (1.8~eV) calculated
within LDA+CDMFT is in good agreement with the
optical conductivity measurements (showing the peak corresponding to vertical transitions
at $\sim$1.8~eV). The approaches which neglect the electron
correlations (LDA), or treat these
correlation in a simplified manner (LDA+U \cite{ldau}), do not provide
such an agreement. Based on LDA+CDMFT calculations,
we establish that Mn$_4$ is a predominantly Mott-Hubbard insulator
with strong electron correlations. These correlations are important
for description of ground-state properties of Mn$_4$
(intra-molecule exchange interactions, spin ground state, etc.)
and may be crucial for future studies of 
transport through Mn$_4$ molecules.

It is important to note that DMFT currently provides the most 
advanced description of the correlation effects, and is
actively used to describe correlations in many materials \cite{dmft}. 
However, its
use so far has been restricted to the systems containing only
a few atoms per unit cell \cite{dmft,mazur,ti2o3,tiocl}. The present work is, to
our knowledge, the first example of applying the LDA+CDMFT
method to large systems with very low molecular symmetry,
containing 80 atoms per unit cell, establishing
feasibility of such calculations and reliability of results.
Moreover, the LDA+CDMFT method allows studies of
the electronic structure of Mn$_4$ at finite temperatures.

The family of Mn$_4$ SMMs \cite{Mn4qtm,Wang1996,mn4exch,mn4chains}
includes different compounds
which have structurally similar cores made of four Mn ions
located in the corners of distorted tetrahedron, but possess
different ligands, and exhibit different magnetic and
electronic properties. The
(pyH)$_3$[Mn$_4$O$_3$Cl$_7$(OAc)$_3$]$\cdot$2MeCN
molecules studied here contain
three ferromagnetically coupled Mn$^{3+}$ ions (spin 2) which interact
antiferromagnetically with the fourth Mn$^{4+}$ ion (spin 3/2),
thus leading to the total ground-state spin $S=9/2$.

%
Electronic structure of many SMMs, including dimers
of the Mn$_4$ family, has been studied previously
using the density functional theory within
generalized gradient approximation (GGA)
\cite{pederson,pedersonMn4}.
These methods describe well the magnetic moments of individual
ions and magnetic anisotropy energy, but do not take into account the
correlation effects, which are important
for many transition metal-oxide systems \cite{tmox}, and for
SMMs (where the transition metal ions are coupled
via oxygens) \cite{v15xes,v15chi,Barbour2006,Boukhvalov2007}.
As a result, the value of the
electronic gap is underestimated, while the magnitude
of the superexchange coupling between magnetic ions is seriously
overestimated (by a factor of 3 for V$_{15}$ SMMs and for Mn$_4$ dimers).
The electron correlations can be treated in
a simplified way by using the LDA+U approach \cite{ldau},
which takes into account the on-site Coulomb repulsion between the electrons
(quantified by the energy $U$) and the intra-atomic Hund exchange
(quantified by the parameter $J$). The LDA+U calculations
\cite{v15xes,v15chi,Barbour2006,Boukhvalov2007}
have shown that
even a simplified account of many-body effects in SMMs leads to
much better agreement between theory and experiment: e.g., for
V$_{15}$, the electronic gap, positions of the bands, as well as the
exchange interaction parameters agree with experiment within
20--30\%.

\begin{figure}
\includegraphics[width=3.2 in]{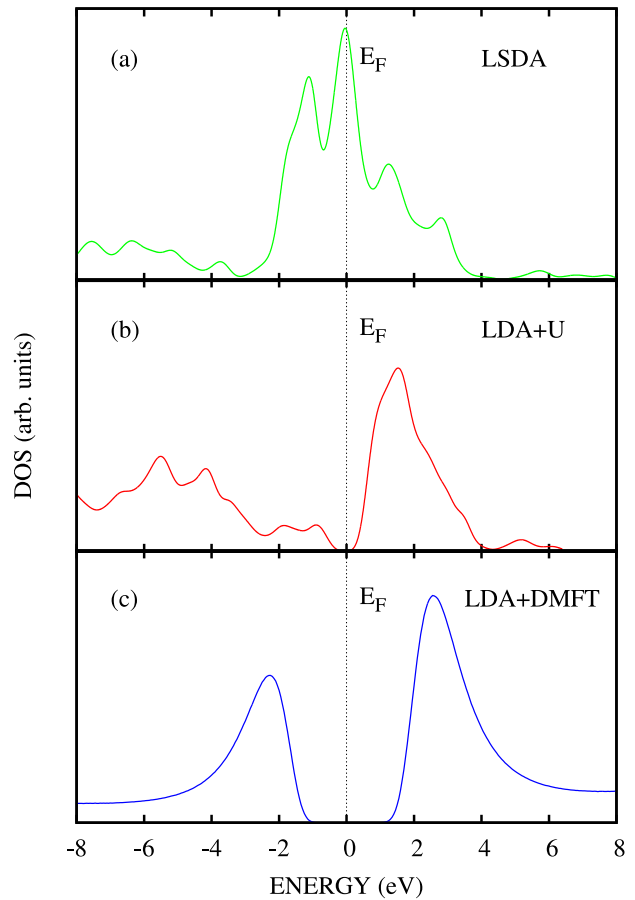}
\caption{\label{Mn4_theory} (Color online). Densities of states (DOS) for
Mn 3d orbitals of Mn$_4$
calculated using: (a) Local Spin Density Approximation (LSDA);
(b) Local Density Approximation taking into account one site Coulomb
repulsin (LDA+U) for $U=4$~eV, $J=0.9$~eV; (c) LDA+CDMFT approach for
$U=4$~eV, $J=0.9$~eV, 
$\beta$ = 8 eV$^{-1}$.}
\end{figure}

For electronic structure of Mn$_4$, 
we first used the single-electron LDA and LDA+U methods. The calculations have
been performed using the ASA-LMTO (Atomic Sphere Approximation ---
Linear Muffin-Tin Orbitals) method \cite{LMTO} as reported in
our earlier works
\cite{v15xes,v15chi,Barbour2006,Boukhvalov2007,mn12mn6film1}.
Based on the agreement with experiments achieved in
Refs.~\cite{Boukhvalov2007,mn12mn6film1},
we used the values $U=4$~eV and $J=0.9$~eV for Mn atoms.
The experimentally determined structure was employed.
The resulting densities of states (DOS)
are given in Fig.~\ref{Mn4_theory} with small additional broadening
to make a simpler comparison with LDA+CDMFT results.
The LDA results, giving
finite electron density at the Fermi energy, can not reproduce the correct
insulating states of SMMs, and
are not discussed.

The LDA+U calculations give a finite gap of 0.9~eV,
thus demonstrating the importance of Coulomb interaction effects
\cite{ldaunote}.
However, the calculated gap is much smaller than the 
experimentally measured one ($\sim 1.8$~eV for vertical excitations, 
see below). Also, LDA+U calculations
give no information about the dependence of the gap on temperature,
and the nature of the
gap is unclear: it may be of band origin, magnetic origin, or
Hubbard origin (caused by Coulomb repulsion between electrons).
Furthermore, the LDA+U scheme predicts correct values for
the magnetic moments, 3.20~$\mu_B$ on Mn$^{4+}$ and 4.32~$\mu_B$ on
Mn$^{3+}$ ions. Also, we calculated the
exchange parameters $J_{ij}$ in the Heisenberg spin Hamiltonian
$H=\sum_{i,j} J_{ij}{\bf S}_i\cdot{\bf S}_j$
(where $S_i$ are the spins of Mn ions), assuming spin configuration
where Mn$^{3+}$ spins are directed up, and the Mn$^{4+}$ spin
is directed down. The calculated
couplings between Mn$^{4+}$ and Mn$^{3+}$ are -29, -14, and -12 K,
and the couplings within the Mn$^{3+}$ triangle are about 75 K,
which is in qualitative agreement with
the previously reported exchanges \cite{Wang1996,mn4exch}.
The exact diagonalization of the Heisenberg exchange Hamiltonian
yields the correct $S=9/2$ ground state of Mn$_4$, but
the excited spin states are not reproduced correctly.
Also, the exchange parameters strongly depend on the spin
configuration: for ferromagnetic arrangement of all Mn spins
(i.e., with reversed spin of Mn$^{4+}$),
the Mn$^{3+}$--Mn$^{4+}$ exchanges increase by a factor of two.
Therefore, the LDA+U scheme provides only a qualitative description
of electronic and magnetic properties of Mn$_4$.
This is in striking contrast with the V$_{15}$ SMM, where
the LDA+U results are in a very good quantitative agreement with a wide range of
experiments, from X-ray spectroscopy to magnetic susceptibility
measurements \cite{v15chi,v15xes}. The 
difference is due to the limited account of the
correlation effects by the LDA+U scheme:
for V$_{15}$, where many-body effects are moderate,
LDA+U is adequate, while in Mn$_4$, where
the many-body effects are very strong, this is not so.

A detailed account of the electron correlations for Mn$_4$ is achieved
by using the dynamical mean-field theory (DMFT) within the
cluster LDA+DMFT scheme \cite{cdmft1,mazur,ti2o3}.
Unlike density functional approaches, DMFT considers
the total energy of the system
(more accurately, the thermodynamic potential) 
as a functional of the Green's function instead of the
density matrix. The analytical properties of the Green's function
guarantee that the knowledge of the
spectral density is equivalent to the knowledge of the time-dependent
Green's function, whereas the density matrix is only static value of the
latter. The CDMFT approach maps the many-body crystal system onto
an effective self-consistent multi-orbital quantum
impurity-cluster problem \cite{cdmft1}.
The Green's function matrix is 
calculated via Brillouin zone integration.
We down-folded the full LMTO Hamiltonian to the basis of four Mn
$d$-orbitals in the unit cell and performed CDMFT calculations with
the four Mn atoms. We found that hybridization to neighboring Mn atoms in SMM crystal
is negligible, and already the first CDMFT iteration gives a reasonable
solution of the LDA+CDMFT scheme. The only important quantity which needs to be
found self-consistently is the value of the many-body chemical potential,
which gives the number of $d$-electrons in the Mn$_4$ cluster.

The cluster impurity solution within LDA+CDMFT method was carried out
using the multiorbital QMC simulations \cite{ti2o3} for 4 Mn atoms with full $d$-shell
basis with the same Coulomb interaction parameters ($U=4$~eV),
starting from high temperature $\beta$ = 8 eV$^{-1}$
(corresponding to the temperature $T=1450$ K)
down to the actual experimental temperature $T=305$ K ($\beta$ = 38 eV$^{-1}$).
The value of the gap given by the LDA+CDMFT method
is 1.80~eV, twice higher than the gap given by the LDA+U method, and
is practically independent of temperature.
The DOS were calculated using maximum-entropy continuation from
the imaginary time axis, the results are given in Fig.~\ref{Mn4_theory}c
for $T=1450$~K \cite{noteMaxEnt}.
One can see the formation of broad Hubbard bands
due to strong interactions between four Mn atoms in the cluster. Note that similar
broad Hubbard bands exist in theoretical description and x-ray spectroscopy of manganite
compounds \cite{wessley}.

%

\begin{figure}
\includegraphics[width=3.2in]{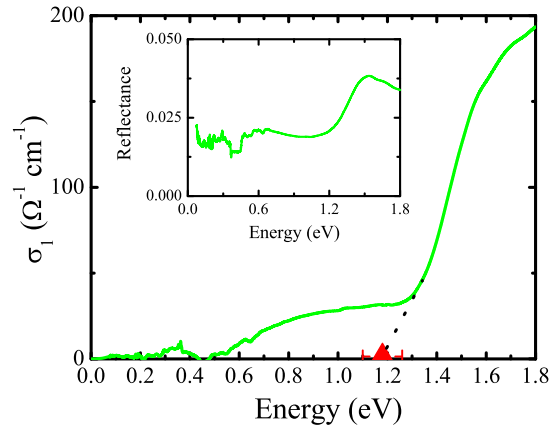}
\caption{\label{Mn4_OpticalGap} (Color online) Optical conductivity
of Mn$_4$ at 300 K,
obtained from a Kramers-Kronig analysis of the measured reflectance
(inset) on the [2~$\bar 2$~1] crystal face. The red triangle shows
the onset of optical aborption due to Mn charge transfer excitations
with the corresponding error bar ($1.18 \pm 0.08$~eV). The peak
($\sim$1.8 eV) corresponds to vertical excitations. The residual AC
conductivity between 0.6 and 1.2 eV is attributable to the organic
constituents. }
\end{figure}

In order to check our LDA+CDMFT calculations,
in particular, the value of the gap (which can clearly differentiate
between the LDA, LDA+U, and LDA+CDMFT results), we measured the optical properties of
Mn$_4$.
Figure \ref{Mn4_OpticalGap} displays the optical conductivity of
Mn$_4$ measured on the
[2~$\bar 2$~1] crystal face \cite{expdetail}.
Based upon the electronic structure results
and comparison with chemically-similar model materials, we assign
the 1.8~eV peak to a superposition of Mn$^{3+}$ to Mn$^{4+}$
charge transfer excitations of the distorted 
[Mn$_4$O$_3$Cl]$^{6+}$ core and (much lower intensity) $d$-to-$d$ on-site excitations of
the Mn centers. Interest in microscopic conduction pathways in
molecular magnets has motivated several combined transport, optical,
and theoretical studies
\cite{Oppenheimer2002,North2003,Baruah2004,Ni2006,Barbour2006,Boukhvalov2007}.
These investigations showed that exact gap values depend on the
nature of the experimental probe and its associated length scale,
although overall trends for chemically-similar materials are
consistent. While the optical gap in
Mn$_4$
is experimentally
determined by extrapolation of the leading edge of the 1.8 eV band
to zero conductivity (1.18 $\pm$0.08 eV), the peak of this
excitation marks the $\Delta$$k$ = 0 transitions and the maximum in
the joint density of states. The peak value ($\sim$1.8 eV) should
thus be compared with theoretical predictions.  The weaker feature
between 0.6~eV and 1.2~eV
derives from the organic constituents, consistent with data from the
[3~2~$\bar 1$] crystal face  that has increased sensitivity to these
structures.

The agreement between experiment and the LDA+CDMFT predictions, 
and lack of such agreement for LDA and LDA+U methods,
clearly demonstrates the importance of the
correlation effects in Mn$_4$ SMMs. These correlations 
may persist in functionalized Mn$_4$ SMMs:
deposition of SMMs on the metal substrates may have little effect
on their electronic structure (unless charging effects or
mechanical deformations occur) \cite{mn12mn6film1,FePorfSubstr}. 
The correlation effects may also be crucial for transport through
Mn$_4$, due to peculiar nature of carriers in Hubbard bands,
and for intra-molecular exchanges (which are still too complex for
LDA+CDMFT).
Our work presents only the first step in this interesting
direction, and further studies are required.

Summarizing, we have investigated the correlation effects
in electronic structure of Mn$_4$ SMMs.
Theoretical studies employed 
(i) the local density approximation, LDA, which
neglects the electron correlations; (ii) LDA+U method,
which describes correlations in a simplified way; and (iii)
LDA+CDMFT method, which provides a detailed account
of the many-electron effects within the cluster dynamical mean-field
theory. The measurements of the
optical conductivity have been used to determine the electronic gap
in (pyH)$_3$[Mn$_4$O$_3$Cl$_7$(OAc)$_3$]$\cdot$2MeCN. 
Among the
three theoretical approaches, only LDA+CDMFT predicts a gap
consistent with experimental measurements, clearly demonstrating
importance of the electron correlations. The
LDA+CDMFT calculations evidence
the predominantly Hubbard insulator state of Mn$_4$.
Based on similarities with other Mn-oxide 
strongly correlated systems \cite{solovyev,dagotto}, one
may expect the correlations to be important for magnetic and
transport properties of Mn$_4$.

Work at the Ames Laboratory was supported by the Department of
Energy --- Basic Energy Sciences under Contract No. DE-AC02-07CH11358.
Work at UT is supported by the Department of Energy 
(subcontract from Ames Laboratory and Contract No. DE-FG02-01ER45885).
Support from FOM (the Netherlands) is acknowledged.


\begin{thebibliography}{99}
\bibitem{ldaunote} Larger $U$ does not improve the LDA+U results.
For $U=5$ eV, the gap is 1.05 eV; to reproduce
the experimental gap of 1.8 eV, $U\sim 8$ eV is needed, which
is unreasonably large and gives unphysical electronic 
structure (incorrect spin of the ground state,
charge transfer spectrum instead of Mott insulator spectrum, etc.).
\bibitem{noteMaxEnt} At lower temperatures, while the value of the gap
can still be reliably determined, the shape of DOS is not dependable:
the QMC errors are too hight to restore a dependable DOS
curve.
\end{thebibliography}
\end{document}